\documentclass[twocolumn,showpacs,preprintnumbers]{revtex4}%
\usepackage{amssymb}
\usepackage{amsmath}
\usepackage{graphicx}
\usepackage{dcolumn}
\usepackage{bm}%
\usepackage{amsfonts}
\begin{document}
\title{Dephasing due to Intermode Coupling in Superconducting Stripline Resonators}
\author{Eyal Buks}
\affiliation{Department of Electrical Engineering, Technion, Haifa 32000 Israel}
\author{Bernard Yurke}
\affiliation{Bell Laboratories, Lucent Technologies, 600 Mountain Avenue, Murray Hill, NJ 07974}
\date{\today }

\begin{abstract}
The nonlinearity exhibited by the kinetic inductance of a superconducting
stripline couples stripline resonator modes together in a manner suitable for
quantum non-demolition measurement of the number of photons in a given
resonator mode. Quantum non-demolition measurement is accomplished by
coherently driving another resonator mode, referred to as the
\textit{detector} mode, and measuring its response. \ We show that the
sensitivity of such a detection scheme is directly related to the dephasing
rate induced by such an intermode coupling. \ We show that high sensitivity is
expected when the detector mode is driven into the nonlinear regime and
operated close to a point where critical slowing down occurs.

\end{abstract}
\maketitle



%It is always \today, today,
%but any date may be explicitly specified

%PACS, the Physics and Astronomy
%Classification Scheme.

\section{Introduction}

The resistive and inductive nonlinearity exhibited by superconducting
striplines and microstrips originates from the nonuniform distribution of the
microwave current over the transmission line cross section. Along the edges,
where the current density obtains its peak value, the current density can
become overcritical, even with relatively moderate power levels. As a result,
the superconducting current distribution may change as the total current
within the transmission line changes. This, in turn, causes both the
inductance $L$ and resistance $R$ per unit length to become current dependent
\cite{Dahm97}.

Such nonlinearity introduces coupling between different modes in a stripline
resonator. \ In the rotating wave approximation (RWA) such coupling adds a
term to the Hamiltonian of the system given by $\sum_{n^{\prime}\neq
n^{\prime\prime}}\hbar\lambda_{n^{\prime}n^{\prime\prime}}N_{n^{\prime}%
}N_{n^{\prime\prime}}$, where $N_{n}=A_{n}^{\dag}A_{n}$ is the number operator
of mode $n$, and the coupling constants $\lambda_{n^{\prime}n^{\prime\prime}}$
are given by \ref{lambda_n'n''}. \ As was shown in Ref. \cite{Sanders 89},
such inter-mode coupling may allow a non-demolition measurement
\cite{braginsky}, \cite{Caves 80} of the number of photons in a
\textit{signal} mode by intensively driving another mode, called a
\textit{detector} mode (or \textit{pump}), and monitoring its response near
the pump frequency. \ Such a measurement scheme is characterized by a
measurement time, defined as the time needed to distinguish between initial
states of the signal mode having different number of photons. \ Single-photon
detection can be realized if the measurement time can be made shorter than the
lifetime of a Fock state of the signal mode.

However, according to the Bohr's complementarity principle \cite{Bohr 49}, in
the limit of single-photon detection, dephasing is expected to come into play,
leading to broadening of the resonance line shape of the signal mode. \ Thus,
a \textit{which-path}-like experiment can be employed to test whether
single-photon detection is possible. \ This can be done by monitoring the
resonance response of the signal mode and the way it is affected by driving
the detector mode. \ Observing a broadening comparable to or larger than the
width of the resonance when the detector is not driven implies that
single-photon detection is possible in principle, namely, the measurement time
is comparable to or shorter than the lifetime of a photon in the mode.

In practice, however, as we show below, the coupling constants $\lambda
_{n^{\prime}n^{\prime\prime}}$ in superconducting stripline resonators are far
too weak to allow single-photon detection when the detector mode is taken to
have a linear response. \ On the other hand, the same kinetic inductance which
leads to intermode coupling also gives rise to a detector mode Kerr
nonlinearity. \ In the RWA such nonlinearity adds the term $\sum_{n}%
K_{n}\left(  N_{n}\right)  ^{2}$ to the Hamiltonian of the system, where the
Kerr constants $K_{n}$ are given by \ref{K_n'}. \ Such Duffing-like
nonlinearity leads to bistability when the drive level exceeds some critical
value. \ In the bistable regime of operation the response of the detector mode
exhibits hysteresis and jumps as the frequency of the drive or its amplitude
is varied.

In the present work we investigate the dephasing rate induced by the driven
detector mode on the signal mode. \ We consider a detector mode having both a
Kerr-like nonlinearity and nonlinear damping. \ We find that the dephasing
rate diverges as one approaches a point where the response of the detector
mode exhibits a jump. \ At such a point the slope of the response vs.
frequency is infinite. \ This result suggests that strong dephasing can be
induced when driving the detector mode in the nonlinear regime, even with
relatively weak intermode coupling. \ Thus, operating such a photon detector
in this regime may allow enhanced sensitivity.

The dynamics of a driven single-mode stripline resonator having a Kerr-like
nonlinearity, in addition to linear and nonlinear damping \cite{yurke84}, was
studied recently by us in \cite{yurke 05}. \ The equations of motion were
derived using the input-output theory of Gardiner and Collett
\cite{gardiner85,Gea90}. \ We have shown that in the appropriate limit such a
system can serve as a phase-sensitive amplifier whose noise performance
exceeds that of the quantum limits imposed on linear, phase-insensitive
parametric amplifiers \cite{movshovich90}. \ Moreover, we have studied
degradation of device performance, due to two-photon absorption originated by
nonlinear resistance, which can be significant in such devices (see also
\cite{gerry93,gilles94,li95,ho95}).

A closely related problem, the non-demolition measurement of Fock states in
mesoscopic mechanical oscillators, was recently studied theoretically in
\cite{Santamore 04 A}. \ Anharmonic effects on such a measurement were further
studied in \cite{Santamore 04 B}. \ In the RWA the Hamiltonian of the
mechanical systems, studied in \cite{Santamore 04 A} and \cite{Santamore 04
B}, and the one of a stripline resonator presented here, are similar.
\ However, our analysis goes beyond that of \cite{Santamore 04 A} and
\cite{Santamore 04 B} by taking into account the effect of nonlinear damping
\cite{zaitsev}. \ Moreover, while the emphasis in Refs. \cite{Santamore 04 A}
and \cite{Santamore 04 B} is on the properties of the detection scheme,
including a thorough consideration of the phase diffusion back-action, here we
focus on the relation between decoherence and the distinguishability between
states with different numbers of photons.

\section{Single-photon Detector}

The Hamiltonian of the system is given by
\begin{equation}
H=H_{s}+H_{D}+V\ .\label{H}%
\end{equation}
Here $H_{s}$ is the Hamiltonian of the mode of the photons to be detected
(signal mode),
\begin{equation}
H_{s}=\hbar\omega_{s}N_{s}\ ,
\end{equation}
where $N_{s}=A_{s}^{\dagger}A_{s}$ is the number operator of the signal mode.
$H_{D}$ is the Hamiltonian of the detector. $V$ is the coupling between the
signal mode and the detector, and is taken to have the form
\begin{equation}
V=\hbar\lambda N_{s}W,
\end{equation}
where $W$ is an operator on the Hilbert space of the detector.

The Heisenberg equation of motion can be used to show that the number of
photons in the signal mode $N_{s}$ is a constant of motion
\begin{equation}
i\hbar\frac{d}{dt}\left\langle N_{s}\right\rangle =\left\langle [N_{s}%
,H]\right\rangle =0.
\end{equation}
Thus, the detection process in this case is a quantum non-demolition
measurement of the number $N_{s}$ \cite{Sanders 89}.

To study the ability of the detector to measure $N_{s}$ we follow the approach
presented in Refs. \cite{Stern 90}, \cite{Levinson 97}, \cite{Buks 98}.
Consider an initial state at time $t=0$, having a superposition of $0$ and $1$
photons
\begin{equation}
\left\vert \psi\left(  t=0\right)  \right\rangle =\left(  \left\vert
0\right\rangle _{s}+\left\vert 1\right\rangle _{s}\right)  \otimes\left\vert
\chi_{i}\right\rangle _{D},
\end{equation}
where $\left\vert n\right\rangle _{s}$ is a Fock state of the signal mode and
$\left\vert \chi_{i}\right\rangle _{D}$ represents an initial state of the
detector. \ At a later time $t>0$ the state in the Schr\"{o}dinger
representation will evolve into
\begin{equation}
\left\vert \psi\left(  t\right)  \right\rangle =u\left(  t\right)  \left\vert
\psi\left(  t=0\right)  \right\rangle ,
\end{equation}
where the time evolution operator $u\left(  t\right)  $ is given by
\begin{equation}
u\left(  t\right)  =\mathcal{T}\exp\left[  -\frac{i}{\hbar}\int_{0}%
^{t}dt^{\prime}H\left(  t^{\prime}\right)  \right]  ,
\end{equation}
and $\mathcal{T}$ is the time-ordering operator. \ Equation \ref{H} yields
\begin{align}
& u\left(  t\right)  \left\vert n\right\rangle _{s}\otimes\left\vert \chi
_{i}\right\rangle _{D}\\
& =\mathcal{T}\exp\left[  -\frac{i}{\hbar}\int_{0}^{t}dt^{\prime}\left[
H_{D}+\hbar\left(  \omega_{s}+\lambda W\right)  N_{s}\right]  \right]
\left\vert n\right\rangle _{s}\otimes\left\vert \chi_{i}\right\rangle
_{D}\nonumber\\
& =\left\vert n\right\rangle _{s}\otimes u_{n}\left\vert \chi_{i}\right\rangle
_{D},\nonumber
\end{align}
where
\begin{equation}
u_{n}=e^{-in\omega_{s}t}\mathcal{T}\exp\left[  -\frac{i}{\hbar}\int_{0}%
^{t}dt^{\prime}\left(  H_{D}+n\hbar\lambda W\right)  \right]  .\label{u_n}%
\end{equation}
Thus, as expected, the time evolution preserves $N_{s}$, and
\begin{equation}
\left\vert \psi\left(  t\right)  \right\rangle =\left\vert 0\right\rangle
_{s}\otimes\left\vert \chi_{0}\right\rangle _{D}+\left\vert 1\right\rangle
_{s}\otimes\left\vert \chi_{1}\right\rangle _{D},
\end{equation}
where $\left\vert \chi_{0}\right\rangle _{D}=u_{0}\left\vert \chi
_{i}\right\rangle _{D}$ and $\left\vert \chi_{1}\right\rangle _{D}%
=u_{1}\left\vert \chi_{i}\right\rangle _{D}$. The distinguishability between 0
and 1 photons associated with this state is characterized by the parameter
\begin{equation}
\nu\equiv\left\vert \left\langle A_{s}\right\rangle \right\vert ^{2}%
=\;\left\vert _{D}\left\langle \chi_{0}|\chi_{1}\right\rangle _{D}\right\vert
^{2}=\;\left\vert _{D}\left\langle \chi_{i}\right\vert u_{0}^{\dag}%
u_{1}\left\vert \chi_{i}\right\rangle _{D}\right\vert ^{2}.
\end{equation}

Our goal is to find the characteristic time scale $\tau_{\varphi}$ over which
$\nu$ decays from its initial value $\nu=1$ at time $t=0$. \ Regarding the
coupling parameter $\lambda$ as small, one can evaluate $\nu$ using
perturbation theory. \ To lowest order in $\lambda$ one finds using
\ref{|<O(t)>|^2}
\begin{equation}
\nu=1-\lambda^{2}%
%TCIMACRO{\tint \limits_{0}^{t}}%
%BeginExpansion
{\textstyle\int\limits_{0}^{t}}
%EndExpansion
dt^{\prime}%
%TCIMACRO{\tint \limits_{0}^{t}}%
%BeginExpansion
{\textstyle\int\limits_{0}^{t}}
%EndExpansion
dt^{\prime\prime}K\left(  t^{\prime},t^{\prime\prime}\right)  ,\label{ni}%
\end{equation}
where
\begin{equation}
\hat{K}\left(  t^{\prime},t^{\prime\prime}\right)  =\frac{1}{2}\left[
\left\langle \tilde{W}\left(  t^{\prime}\right)  \tilde{W}\left(
t^{\prime\prime}\right)  \right\rangle +\left\langle \tilde{W}\left(
t^{\prime\prime}\right)  \tilde{W}\left(  t^{\prime}\right)  \right\rangle
\right]  ,\label{K(t',t'')}%
\end{equation}
and
\begin{equation}
\tilde{W}\left(  t\right)  =W\left(  t\right)  -\left\langle W\left(
t\right)  \right\rangle .
\end{equation}
For a steady state the correlation function $\hat{K}\left(  t^{\prime
},t^{\prime\prime}\right)  $ is a function of $\tau=\left\vert t^{\prime
}-t^{\prime\prime}\right\vert $. \ Moreover, it decays on some characteristic
time scale $\tau_{\varphi}$, which can be identified from Eq. \ref{ni}
\cite{Levinson 97}
\begin{equation}
\frac{1}{\tau_{\varphi}}=\lambda^{2}\int_{-\infty}^{\infty}d\tau K\left(
\tau\right)  ,\label{tau_phi}%
\end{equation}
where $K\left(  \tau\right)  =\hat{K}\left(  0,\tau\right)  $.

\section{The Detector Mode}

The detector mode is a resonator mode distinct from the signal mode. Because
it is driven to large amplitude, the Kerr nonlinearity and the linear and
nonlinear damping of the mode are taken into account in our analysis. \ Here
we review the main results of Ref. \cite{yurke 05}.

\subsection{The Hamiltonian}

The Hamiltonian is given by%
\begin{equation}
H_{D}=H_{A}+H_{K}+H_{a_{1}}+H_{a_{2}}+H_{a_{3}}+H_{T_{1}}+H_{T_{2}}+H_{T_{3}%
}\ ,
\end{equation}
where $H_{A}$ is the Hamiltonian for the detector mode
\begin{equation}
H_{A}=\hbar\omega_{0}A^{\dagger}A\ ,
\end{equation}
$H_{K}$ is the Hamiltonian for the Kerr nonlinearity
\begin{equation}
H_{K}=\frac{\hbar}{2}KA^{\dagger}A^{\dagger}AA\ ,
\end{equation}
$H_{a1}$, $H_{a2}$ and $H_{a3}$ are the Hamiltonians for the bath modes
associated with the dissipative elements
\begin{equation}
H_{a1}=\int d\omega\hbar\omega a_{1}^{\dagger}(\omega)a_{1}(\omega)\ ,
\end{equation}%
\begin{equation}
H_{a2}=\int d\omega\hbar\omega a_{2}^{\dagger}(\omega)a_{2}(\omega)\ ,
\end{equation}%
\begin{equation}
H_{a3}=\int d\omega\hbar\omega a_{3}^{\dagger}(\omega)a_{3}(\omega)\ .
\end{equation}
$H_{T_{1}}$ and $H_{T_{2}}$ are the Hamiltonians for the linear dissipative
elements, one of which is the port through which signals enter and leave the
oscillator. These Hamiltonians linearly couple the bath modes, $a_{1}$ and
$a_{2}$ respectively, to the oscillator mode $A$
\begin{equation}
H_{T_{1}}=\hbar\int d\omega\lbrack T_{1}A^{\dagger}a_{1}(\omega)+T_{1}^{\ast
}a_{1}^{\dagger}(\omega)A]\ ,
\end{equation}%
\begin{equation}
H_{T_{2}}=\hbar\int d\omega\lbrack T_{2}A^{\dagger}a_{2}(\omega)+T_{2}^{\ast
}a_{2}^{\dagger}(\omega)A]\ .
\end{equation}
The two-photon absorptive coupling of the resonator mode to the bath modes
$a_{3}$ is modeled by a hopping Hamiltonian in which two cavity photons are
destroyed for every bath photon created \cite{tornau74, agarwal86, gilles93,
ezaki99, kitamura99}
\begin{equation}
H_{T_{3}}=\hbar\int d\omega\lbrack T_{3}A^{\dagger}A^{\dagger}a_{3}%
(\omega)+T_{3}^{\ast}a_{3}^{\dagger}(\omega)AA].
\end{equation}
The modes are boson modes, satisfying the usual Bose commutation relations.

\subsection{Equations of motion for $A$}

The equation of motion for $A$ is given by%

\begin{align}
\frac{dA}{dt}  & =-i\omega_{0}A-iKA^{\dagger}AA-\gamma A-\gamma_{3}A^{\dagger
}AA\label{dA/dt}\\
& -i\sqrt{2\gamma_{1}}e^{i\phi_{1}}a_{1}^{in}(t)-i\sqrt{2\gamma_{2}}%
e^{i\phi_{2}}a_{2}^{in}(t)\nonumber\\
& -i2\sqrt{\gamma_{3}}e^{i\phi_{3}}A^{\dagger}a_{3}^{in}(t),\nonumber
\end{align}
where $\gamma=\gamma_{1}+\gamma_{2}$. \ The damping constants are written as%
\begin{equation}
T_{n}=\sqrt{\frac{\gamma_{n}}{\pi}}e^{i\phi_{n}},
\end{equation}
where $\gamma_{n}$ and $\phi_{n}$ $(n=1,2,3)$ are real, and
\begin{equation}
a_{n}^{in}(t)=\frac{1}{\sqrt{2\pi}}\int d\omega e^{-i\omega(t-t_{0})}%
a_{n}(t_{0},\omega).
\end{equation}
The following commutation relations exist for the bath modes%
\begin{align}
\lbrack a_{n}(t_{0},\omega),a_{m}^{\dagger}(t_{0},\omega^{\prime})]  &
=\delta(\omega-\omega^{\prime})\delta_{nm} \ ,\\
\ [a_{n}(t_{0},\omega),a_{m}(t_{0},\omega^{\prime})]  & =0 \ ,
\end{align}
where $n,m=1,2,3$.

\subsection{Mean-Field Solution}

Since the detector mode is driven to large amplitude, compared with quantum
fluctuations, the behavior of the mean-field is the classical behavior of the
mode, except very near instability points. Hence, to obtain expressions for
the mean-field behavior, in this section, we treat $a_{1}^{in}$, $a_{2}^{in}$,
and $a_{3}^{in}$ as complex numbers rather than as operators. We take the bath
mode amplitudes $a_{2}^{in}$ and $a_{3}^{in}$ to be those entering the cavity
from losses. Consequently, we take $a_{2}^{in}=0$ and $a_{3}^{in}=0$. The
amplitude $a_{1}^{in}$ is chosen to be that of the incoming drive and is taken
to have the oscillatory time dependence
\begin{equation}
a_{1}^{in}=b_{1}^{in}e^{-i(\omega_{p}t+\psi_{1})},
\end{equation}
where $b_{1}^{in}$ is a real constant. Writing $A$ as
\begin{equation}
A=Be^{-i(\omega_{p}t+\phi_{B})},
\end{equation}
where $B$ is a positive real constant, the equations of motion become
\begin{equation}
\lbrack i(\omega_{0}-\omega_{p})+\gamma]B+(iK+\gamma_{3})B^{3}=-i\sqrt
{2\gamma_{1}}b_{1}^{in}e^{i(\phi_{1}+\phi_{B}-\psi_{1})}\ .\label{mfs}%
\end{equation}
Multiplying each side of the nonlinear equation by its complex conjugate, one
obtains \cite{nayfeh}, \cite{landau}
\begin{gather}
B^{6}+\frac{2[(\omega_{0}-\omega_{p})K+\gamma\gamma_{3}]}{K^{2}+\gamma_{3}%
^{2}}B^{4}\label{cubic_eq_for_E}\\
+\frac{(\omega_{0}-\omega_{p})^{2}+\gamma^{2}}{K^{2}+\gamma_{3}^{2}}%
B^{2}-\frac{2\gamma_{1}}{K^{2}+\gamma_{3}^{2}}(b_{1}^{in})^{2}=0\ .\nonumber
\end{gather}
Once $B^{2}$ has been determined from the above cubic algebraic equation, the
phase can be found from%
\begin{equation}
\cot(\phi_{1}+\phi_{B}-\psi_{1}-\pi/2)=\frac{\gamma+\gamma_{3}B^{2}}%
{\omega_{0}-\omega_{p}+KB^{2}}.
\end{equation}
Taking the derivative of Eq.~\ref{cubic_eq_for_E} with respect to $\omega_{p}%
$, one finds
\begin{align}
& \frac{\partial B^{2}}{\partial\omega_{p}}\label{dE/d(w_p)}\\
& =\frac{2(\omega_{0}-\omega_{p}+KB^{2})B^{2}}{(\omega_{0}-\omega_{p}%
+2KB^{2})^{2}+\left(  \gamma+2\gamma_{3}B^{2}\right)  ^{2}-(K^{2}+\gamma
_{3}^{2})B^{4}}.\nonumber
\end{align}

\subsection{Onset of bistability point}

At the onset of bistability point the following holds%
\begin{equation}
\frac{\partial\omega_{p}}{\partial B^{2}}=\frac{\partial^{2}\omega_{p}%
}{\partial\left(  B^{2}\right)  ^{2}}=0.
\end{equation}
Such a point occurs only if the nonlinear damping is sufficiently small%
\begin{equation}
|K|>\sqrt{3}\gamma_{3}.
\end{equation}
At this critical point the following holds%
\begin{equation}
B_{c}^{2}=\frac{2\gamma}{\sqrt{3}(|K|-\sqrt{3}\gamma_{3})}\ ,\label{B^2_c}%
\end{equation}%
\begin{equation}
\omega_{0}-\omega_{pc}=-\gamma\frac{K}{|K|}\left[  \frac{4\gamma_{3}%
|K|+\sqrt{3}(K^{2}+\gamma_{3}^{2})}{K^{2}-3\gamma_{3}^{2}}\right]
,\label{omega_c}%
\end{equation}%
\begin{equation}
(b_{1c}^{in})^{2}=\frac{4}{3\sqrt{3}}\frac{\gamma^{3}(K^{2}+\gamma_{3}^{2}%
)}{\gamma_{1}(|K|-\sqrt{3}\gamma_{3})^{3}}\ .\label{b^in_1c}%
\end{equation}

\subsection{Quantum fluctuations about the mean-field solution}

To determine the quantum fluctuations about the mean-field solution we write
\begin{equation}
a_{1}^{in}=b_{1}^{in}e^{-i(\omega_{p}t+\psi_{1})}+c_{1}^{in}e^{-i\omega_{p}t},
\end{equation}%
\begin{equation}
a_{2}^{in}=c_{2}^{in}e^{-i\omega_{p}t},
\end{equation}%
\begin{equation}
a_{3}^{in}=c_{3}^{in}e^{-i\omega_{p}t},
\end{equation}
and
\begin{equation}
A=Be^{-i(\omega_{p}t+\phi_{B})}+ae^{-i\omega_{p}t},\label{A=B+a}%
\end{equation}
where $B$ constitutes the mean-field amplitude of the detector mode in
response to the classical drive $b_{1}^{in}$. $\ $The operators $c_{1}^{in}$,
$c_{2}^{in}$, $c_{3}^{in}$, and $a$ are regarded as small and will be kept
only up to linear order. \ In this approximation the operator $a$ satisfies
the following second-order equation of motion
\begin{equation}
\frac{d^{2}a}{dt^{2}}+2\Re(w)\frac{da}{dt}+(|w|^{2}-|v|^{2})a=\Gamma\left(
t\right)  ,\label{d^2a/dt^2}%
\end{equation}
where
\begin{equation}
w=i(\omega_{0}-\omega_{p})+\gamma+2(iK+\gamma_{3})B^{2},\label{W}%
\end{equation}%
\begin{equation}
v=(iK+\gamma_{3})B^{2}e^{-2i\phi_{B}},\label{V}%
\end{equation}%
\begin{equation}
\Gamma\left(  t\right)  =\frac{dF\left(  t\right)  }{dt}+w^{\ast
}F(t)-vF^{\dagger}(t),
\end{equation}
and
\begin{align}
F\left(  t\right)   & =-i\sqrt{2\gamma_{1}}e^{i\phi_{1}}c_{1}^{in}%
-i\sqrt{2\gamma_{2}}e^{i\phi_{2}}c_{2}^{in}\\
& -i2\sqrt{\gamma_{3}}Be^{i(\omega_{p}t+\phi_{B}+\phi_{3})}c_{3}%
^{in}\ .\nonumber
\end{align}
The solution to Eq.~\ref{d^2a/dt^2} is given by
\begin{equation}
a(t)=\int_{-\infty}^{\infty}d\tau G(t-\tau)\Gamma(\tau)\ ,\label{a(t)}%
\end{equation}
where
\begin{equation}
G(t)=u(t)\frac{e^{-\lambda_{0}t}-e^{-\lambda_{1}t}}{\lambda_{0}-\lambda_{1}%
},\label{G(t)}%
\end{equation}
$u(t)$ is the step function, and the eigenvalues $\lambda_{0}$ and
$\lambda_{1}$ satisfy
\begin{equation}
\lambda_{0}+\lambda_{1}=2\Re(w),\label{lam_0*+am_1}%
\end{equation}%
\begin{equation}
\lambda_{0}\lambda_{1}=|w|^{2}-|v|^{2}\ .\label{lam_0*lam_1}%
\end{equation}
From Eq. \ref{dE/d(w_p)} one finds that the slope of the response function
$B^{2}$ vs. $\omega_{p}$ is infinite when
\begin{equation}
|w|^{2}-|v|^{2}=0.
\end{equation}
Thus, using Eq. \ref{lam_0*lam_1}, one finds that at these points at least one
of the eigenvalues $\lambda_{0}$ and $\lambda_{1}$ vanishes. The system
eigenmode corresponding to the eigenvalue that vanishes is the one that
experiences critical slowing down.

\subsection{Thermal noise}

Assuming the bath modes are in thermal equilibrium, the following hold
\begin{equation}
\left\langle c_{n}^{in}(\tau)\right\rangle =\left\langle c_{n}^{in\dag}%
(\tau)\right\rangle =0,\label{<c>,<c+>}%
\end{equation}
\begin{equation}
\left\langle c_{n}^{in}(\tau)c_{m}^{in}(\tau^{\prime})\right\rangle
=\left\langle c_{n}^{in\dag}(\tau)c_{m}^{in\dag}(\tau^{\prime})\right\rangle
=0,\label{<cc>,<c+c+>}%
\end{equation}
\begin{equation}
\left\langle c_{n}^{in\dag}(\tau)c_{m}^{in}(\tau^{\prime})\right\rangle
=\delta\left(  \tau-\tau^{\prime}\right)  \delta_{nm}\left\langle
n_{\omega_{0}}\right\rangle ,\label{<c+c>}%
\end{equation}
\begin{equation}
\left\langle c_{n}^{in}(\tau)c_{m}^{in\dag}(\tau^{\prime})\right\rangle
=\delta\left(  \tau-\tau^{\prime}\right)  \delta_{nm}\left(  \left\langle
n_{\omega_{0}}\right\rangle +1\right)  ,\label{<cc+>}%
\end{equation}
where
\begin{equation}
\left\langle n_{\omega}\right\rangle =\frac{1}{e^{\beta\hbar\omega}-1},
\end{equation}
$n,m=1,2,3$ and $\beta=1/k_{B}T$.

\section{The Dephasing Rate}

Since the interaction Hamiltonian between modes is bilinear in the number
operators for the modes, one has $W=A^{\dag}A$, or using Eq.~\ref{A=B+a},
\begin{equation}
W\left(  t\right)  =B^{2}+B\left[  a\left(  t\right)  e^{i\phi_{B}}+a^{\dag
}\left(  t\right)  e^{-i\phi_{B}}\right]  +a^{\dag}\left(  t\right)  a\left(
t\right)  .
\end{equation}
Using Eqs.~\ref{a(t)} and \ref{<c>,<c+>}, one finds
\begin{equation}
\left\langle a\left(  t\right)  \right\rangle =\left\langle a^{\dag}\left(
t\right)  \right\rangle =0.
\end{equation}
Moreover, keeping terms only up to first order in $a$, one finds
\begin{equation}
\tilde{W}\left(  t\right)  =W\left(  t\right)  -\left\langle W\left(
t\right)  \right\rangle =B\left[  a\left(  t\right)  e^{i\phi_{B}}+a^{\dag
}\left(  t\right)  e^{-i\phi_{B}}\right]  .\label{W_D}%
\end{equation}
In order to use Eq. \ref{tau_phi} to calculate the dephasing rate, some
expectation values must be evaluated. \ Using Eqs.~\ref{<c>,<c+>},
\ref{<cc>,<c+c+>}, \ref{<c+c>}, and \ref{<cc+>}, one finds
\begin{equation}
\left\langle F\left(  \tau\right)  F\left(  \tau^{\prime}\right)
\right\rangle =0,
\end{equation}%
\begin{equation}
\left\langle F^{\dagger}\left(  \tau\right)  F^{\dagger}\left(  \tau^{\prime
}\right)  \right\rangle =0,
\end{equation}%
\begin{equation}
\left\langle F\left(  \tau\right)  F^{\dagger}\left(  \tau^{\prime}\right)
\right\rangle =\left(  \lambda_{0}+\lambda_{1}\right)  \delta\left(  \tau
-\tau^{\prime}\right)  \left\langle n_{\omega_{0}}\right\rangle ,\nonumber
\end{equation}
and
\begin{equation}
\left\langle F^{\dagger}\left(  \tau\right)  F\left(  \tau^{\prime}\right)
\right\rangle =\left(  \lambda_{0}+\lambda_{1}\right)  \delta\left(  \tau
-\tau^{\prime}\right)  \left(  \left\langle n_{\omega_{0}}\right\rangle
+1\right)  .
\end{equation}
Thus, using Eq.~\ref{a(t)}, one finds
\begin{align}
& \left\langle a\left(  t\right)  a\left(  t^{\prime}\right)  \right\rangle \\
& =\int_{-\infty}^{\infty}d\tau\int_{-\infty}^{\infty}d\tau^{\prime}%
G(t-\tau)G(t^{\prime}-\tau^{\prime})\left\langle \Gamma(\tau)\Gamma
(\tau^{\prime})\right\rangle \nonumber\\
& =\left(  \lambda_{0}+\lambda_{1}\right)  \left\langle n_{\omega_{0}%
}\right\rangle v\int_{-\infty}^{\infty}d\tau\frac{dG(t-\tau)}{d\tau
}G(t^{\prime}-\tau)\nonumber\\
& +\left(  \lambda_{0}+\lambda_{1}\right)  \left(  \left\langle n_{\omega_{0}%
}\right\rangle +1\right)  v\int_{-\infty}^{\infty}d\tau G(t-\tau
)\frac{dG(t^{\prime}-\tau)}{d\tau}\nonumber\\
& -\left(  \lambda_{0}+\lambda_{1}\right)  \left(  2\left\langle n_{\omega
_{0}}\right\rangle +1\right)  vw^{\ast}\int_{-\infty}^{\infty}d\tau
G(t-\tau)G(t^{\prime}-\tau)\nonumber
\end{align}
and
\begin{align}
& \left\langle a^{\dagger}\left(  t\right)  a\left(  t^{\prime}\right)
\right\rangle \\
& =\int_{-\infty}^{\infty}d\tau\int_{-\infty}^{\infty}d\tau^{\prime}%
G(t-\tau)G(t^{\prime}-\tau^{\prime})\left\langle \Gamma^{\dagger}(\tau
)\Gamma(\tau^{\prime})\right\rangle \nonumber\\
& =\left(  \lambda_{0}+\lambda_{1}\right)  \left(  \left\langle n_{\omega_{0}%
}\right\rangle +1\right)  \int_{-\infty}^{\infty}d\tau\frac{dG(t-\tau)}{d\tau
}\frac{dG(t^{\prime}-\tau)}{d\tau}\nonumber\\
& -\left(  \lambda_{0}+\lambda_{1}\right)  \left(  \left\langle n_{\omega_{0}%
}\right\rangle +1\right)  w^{\ast}\int_{-\infty}^{\infty}d\tau\frac
{dG(t-\tau)}{d\tau}G(t^{\prime}-\tau)\nonumber\\
& -\left(  \lambda_{0}+\lambda_{1}\right)  \left(  \left\langle n_{\omega_{0}%
}\right\rangle +1\right)  w\int_{-\infty}^{\infty}d\tau G(t-\tau
)\frac{dG(t^{\prime}-\tau)}{d\tau}\nonumber\\
& +\left(  \lambda_{0}+\lambda_{1}\right)  \left(  \left\langle n_{\omega_{0}%
}\right\rangle +1\right)  \left\vert w\right\vert ^{2}\int_{-\infty}^{\infty
}d\tau G(t-\tau)G(t^{\prime}-\tau)\nonumber\\
& +\left(  \lambda_{0}+\lambda_{1}\right)  \left\langle n_{\omega_{0}%
}\right\rangle \left\vert v\right\vert ^{2}\int_{-\infty}^{\infty}d\tau
G(t-\tau)G(t^{\prime}-\tau).\nonumber
\end{align}
Using Eq.~\ref{K(t',t'')}, these results yield
\begin{align}
& \frac{\hat{K}\left(  t,t^{\prime}\right)  }{B^{2}\left(  \lambda_{0}%
+\lambda_{1}\right)  \coth\frac{\beta\hbar\omega_{0}}{2}}\\
& =\int_{-\infty}^{\infty}d\tau\frac{dG(t-\tau)}{d\tau}\frac{dG(t^{\prime
}-\tau)}{d\tau}\nonumber\\
& -\operatorname{Re}\left(  w-ve^{2i\phi_{B}}\right) \nonumber\\
& \times\int_{-\infty}^{\infty}d\tau\left[  G(t-\tau)\frac{dG(t^{\prime}%
-\tau)}{d\tau}+\frac{dG(t-\tau)}{d\tau}G(t^{\prime}-\tau)\right] \nonumber\\
& +\left\vert w-ve^{2i\phi_{B}}\right\vert ^{2}\int_{-\infty}^{\infty}d\tau
G(t-\tau)G(t^{\prime}-\tau).\nonumber
\end{align}
From this, by noting that, from Eqs.~\ref{W}, \ref{V} and \ref{mfs},
\begin{equation}
\left(  w-ve^{2i\phi_{B}}\right)  B=-i\sqrt{2\gamma_{1}}b_{1}^{in}%
e^{i(\phi_{1}+\phi_{B}-\psi_{1})}.\label{W-V}%
\end{equation}
and that
\begin{equation}
\int_{-\infty}^{\infty}dt^{\prime}\int_{-\infty}^{\infty}d\tau G(-\tau
)G(t^{\prime}-\tau)=\frac{1}{\left(  \lambda_{0}\lambda_{1}\right)  ^{2}%
},\nonumber
\end{equation}
one obtains, from Eq.~\ref{tau_phi}, the dephasing rate
\begin{equation}
\frac{1}{\tau_{\varphi}}=\lambda^{2}\frac{\lambda_{0}+\lambda_{1}}{\left(
\lambda_{0}\lambda_{1}\right)  ^{2}}2\gamma_{1}\left(  b_{1}^{in}\right)
^{2}\coth\frac{\beta\hbar\omega_{0}}{2}.\label{1/tau_phi}%
\end{equation}
Thus, $1/\tau_{\varphi}$ is directly related to the eigenvalues $\lambda_{0}$
and $\lambda_{1}$ which characterizing the response of the detector mode to
small perturbation. \ As was shown above, the product $\lambda_{0}\lambda_{1}$
vanishes at the points where the slope $\partial B^{2}/\partial\omega_{p}$ is
infinite, leading to a diverging dephasing rate $1/\tau_{\varphi}$. \ However,
our model, which takes nonlinearity into account only to lowest order, breaks
down near these points, where the fluctuation in $A^{\dag}A$ becomes
appreciable. \ To evaluate the actual dephasing rate near these points one has
to take into account higher-order terms in the nonlinear expansion. \ This is
beyond the scope of the present work, although we note that the divergence of
$1/\tau_{\varphi}$ in the present model indicates that the actual dephasing
rate will be relatively large near these points.

\begin{figure}
[ptb]
\begin{center}
\includegraphics[
height=2.751in,
width=3.2283in
]%
{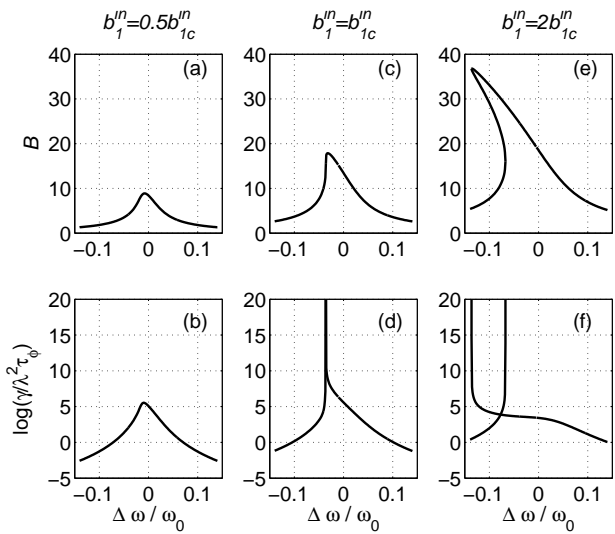}%
\caption{Square root of average photon number in the detector mode $B$ and log
of the normalized dephasing rate $\log\left(  \gamma/\lambda^{2}\tau_{\varphi
}\right)  $, for subcritical drive $b_{1}^{in}=0.5b_{1c}^{in}$ (panels (a) and
(b)), critical drive $b_{1}^{in}=b_{1c}^{in}$ (panels (c) and (d)), and
overcritical drive $b_{1}^{in}=2b_{1c}^{in}$ (panels (e) and (f)). \ The
parameters in this example are $K=-10^{-4}\omega_{0}$, $\gamma_{1}%
=10^{-2}\omega_{0}$, $\gamma_{2}=1.1\gamma_{1}$, and $\gamma_{3}%
=10^{-2}K/\sqrt{3}$.}%
\label{log(deph_rate)}%
\end{center}
\end{figure}

For the linear case $K=0$, $\gamma_{3}=0$ the dephasing rate is given by
\begin{equation}
\frac{1}{\tau_{\varphi}}=\frac{4\lambda^{2}\gamma\gamma_{1}\left(  b_{1}%
^{in}\right)  ^{2}}{\left[  (\omega_{0}-\omega_{p})^{2}+\gamma^{2}\right]
^{2}}\coth\frac{\beta\hbar\omega_{0}}{2}.\label{deph rate linear}%
\end{equation}
Figure \ref{log(deph_rate)} shows $B$, found from solving
Eq.~\ref{cubic_eq_for_E}, and the log of the normalized dephasing rate,
$\log\left(  \gamma/\lambda^{2}\tau_{\varphi}\right)  $, calculated by using
Eq. \ref{1/tau_phi}. \ Three cases are shown, subcritical drive $b_{1}%
^{in}=0.5b_{1c}^{in}$ (panels (a) and (b)), critical drive $b_{1}^{in}%
=b_{1c}^{in}$ (panels (c) and (d)), and overcritical drive $b_{1}^{in}%
=2b_{1c}^{in}$ (panels (e) and (f)). \ The values chosen for this example for
the parameters $K$, $\gamma_{1}$, $\gamma_{2}$, and $\gamma_{3}$ are indicated
in the figure caption. \ As can be seen from Fig. \ref{log(deph_rate)} (d),
the dephasing rate is an asymmetric function of frequency around the onset of
bistability point, where $1/\tau_{\varphi}$ diverges. \ Also, comparing the
two jump points in Fig. \ref{log(deph_rate)} (f) indicates that the divergence
in the right one (with higher frequency) is steeper.%

%TCIMACRO{\FRAME{ftbpFU}{3.2283in}{2.751in}{0pt}{\Qcb{Square root of average
%photon number in the detector mode $B$ and log of the normalized dephasing
%rate $\log\left(  \gamma/\lambda^{2}\tau_{\varphi}\right)  $, for subcritical
%drive $b_{1}^{in}=0.5b_{1c}^{in}$ (panels (a) and (b)), critical drive
%$b_{1}^{in}=b_{1c}^{in}$ (panels (c) and (d)), and overcritical drive
%$b_{1}^{in}=2b_{1c}^{in}$ (panels (e) and (f)). \ The parameters in this
%example are $K=-10^{-4}\omega_{0}$, $\gamma_{1}=10^{-2}\omega_{0}$,
%$\gamma_{2}=1.1\gamma_{1}$, and $\gamma_{3}=10^{-2}K/\sqrt{3}$.}%
%}{\Qlb{log(deph_rate)}}{fig1.eps}{\special{ language "Scientific Word";
%type "GRAPHIC";  maintain-aspect-ratio TRUE;  display "USEDEF";
%valid_file "F";  width 3.2283in;  height 2.751in;  depth 0pt;
%original-width 6.819in;  original-height 5.7718in;  cropleft "0";
%filename 'Fig1.eps';file-properties "XNPEU";}}}%
%BeginExpansion
%EndExpansion

\section{Discussion}

The sensitivity of single-photon detection can be characterized by the
dimensionless parameter $\xi$, defined as
\begin{equation}
\xi\equiv\frac{1}{\gamma\tau_{\varphi}}.
\end{equation}
Consider first the linear case $K=0$, $\gamma_{3}=0$, where the dephasing rate
is given by Eq. \ref{deph rate linear}. \ We assume for simplicity
$2\gamma_{1}\simeq\gamma$, and consider the case of low temperature
$\beta\hbar\omega_{0}\gg1$. \ At resonance where the dephasing rate is
largest, one finds
\begin{equation}
\xi=2B^{2}\left(  \frac{\lambda}{\gamma}\right)  ^{2}.
\end{equation}

In the present case of linear response we assume $B<B_{c}$, where $B_{c}$ is
given by Eq. \ref{B^2_c}. \ Moreover, as can be seen from
Eq.~(\ref{lambda_n'n''}) and (\ref{K_n'}), in many cases $K$ is of the same
order as $\lambda$. \ Thus, assuming also the case where $\gamma_{3}<<K$, one
finds that the largest possible value of $\xi$ in this regime is roughly given
by
\begin{equation}
\xi_{\max}\simeq\frac{\lambda}{\gamma}.
\end{equation}
The experimental data in Refs. \cite{abdo 05A} and \cite{abdo 05B}, together
with Eq. \ref{B^2_c}, allow one to estimate the value of $K$. \ Assuming, as
before, that $\lambda\simeq|K|$, one finds for stripline resonators made of Nb
that $\xi_{\max}\simeq10^{-12}$. \ For resonators made of NbN, on the other
hand, the onset of nonlinear instability occurs at a much lower input power.
\ However, the nature of the nonlinearity observed in NbN resonators is
qualitatively different from what is expected from nonlinearity originated by
the kinetic inductance effect, thus indicating that the underlying physics is
different \cite{abdo 05B}. \ If, however, in spite of this discrepancy, we
employ Eq. \ref{B^2_c} also for the case of NbN resonators, we obtain
$\xi_{\max}\simeq10^{-6}$. We thus find that for both cases $\xi_{\max}$ is
far too small to allow single-photon detection when the pump mode is operated
in the regime of linear response. \ In contrast, as was mentioned above,
operating the pump mode in the regime of nonlinear response may allow a
significant enhancement in the sensitivity. \ Further study is required,
however, to analyze the behavior of the system close to the points where our
model yields divergence in $1/\tau_{\varphi}$.

\appendix

\section{Lossless Transmission Line Resonator}

Consider a lossless linear transmission line with length $l$ extending along
the $x$-axis \cite{yurke84}. \ Let $q\left(  x,t\right)  $ be the charge
density per unit length and define
\begin{equation}
Q\left(  x,t\right)  =\int\limits_{x}^{\infty}dx^{\prime}q\left(  x^{\prime
},t\right)  .
\end{equation}
Thus, $q=-\partial Q/\partial x$ and the voltage across the transmission line
is given by
\begin{equation}
V\left(  x,t\right)  =-\frac{1}{C}\frac{\partial Q}{\partial x},
\end{equation}
where $C$ is the capacitance per unit length along the transmission line. The
current is given by
\begin{equation}
I\left(  x,t\right)  =\frac{\partial Q}{\partial t}.
\end{equation}
The Lagrangian of the system is given by
\begin{align}
\mathcal{L}  & =\frac{1}{2}\int\limits_{0}^{l}dx\left[  LI^{2}-CV^{2}\right]
\\
& =\frac{1}{2}\int\limits_{0}^{l}dx\left[  L\left(  \frac{\partial Q}{\partial
t}\right)  ^{2}-\frac{1}{C}\left(  \frac{\partial Q}{\partial x}\right)
^{2}\right]  ,\nonumber
\end{align}
where $L$ is the inductance per unit length along the transmission line.
\ Here we assume that both $C$ and $L$ depend on $x$. \ Moreover, the kinetic
inductance leads to a dependence of $L$ on the current, given by
\cite{Dahm97},
\begin{equation}
L=L_{0}+\Delta L\left(  \frac{I}{I_{c}}\right)  ^{2}.
\end{equation}
As a basis for expanding $Q\left(  x,t\right)  $ as
\begin{equation}
Q\left(  x,t\right)  =\sum_{n}q_{n}\left(  t\right)  u_{n}\left(  x\right)  ,
\end{equation}
we use the solutions of the boundary value problem
\begin{equation}
\frac{d}{dx}\left(  \frac{1}{C}\frac{du_{n}}{dx}\right)  =-\omega_{n}%
^{2}Lu_{n}\label{we u}%
\end{equation}
with the boundary conditions of vanishing current $u_{n}\left(  0\right)
=u_{n}\left(  l\right)  =0$. \ We assume that the functions $u_{n}\left(
x\right)  $ are chosen to be real. \ Using this basis and the Strum -
Liouville theorem, one obtains
\begin{equation}
\mathcal{L}=\frac{1}{2}\sum_{n}\left(  \dot{q}_{n}^{2}-\omega_{n}^{2}q_{n}%
^{2}\right)  +\Delta\mathcal{L},
\end{equation}
where
\begin{align}
\Delta\mathcal{L}  & =\frac{1}{2I_{c}^{2}}\sum_{n^{\prime},n^{\prime\prime
},n^{\prime\prime\prime},n^{\prime\prime\prime\prime}}\dot{q}_{n^{\prime}}%
\dot{q}_{n^{\prime\prime}}\dot{q}_{n^{\prime\prime\prime}}\dot{q}%
_{n^{\prime\prime\prime\prime}}\\
& \times\int\limits_{0}^{l}dx\Delta Lu_{n^{\prime}}u_{n^{\prime\prime}%
}u_{n^{\prime\prime\prime}}u_{n^{\prime\prime\prime\prime}}.\nonumber
\end{align}
The variable canonically conjugate to $q_{n}$ is
\begin{equation}
p_{n}=\frac{\partial\mathcal{L}}{\partial\dot{q}_{n}}=\dot{q}_{n}%
+\frac{\partial\Delta\mathcal{L}}{\partial\dot{q}_{n}}.\label{can conj}%
\end{equation}
To first order in $\Delta L$ the Hamiltonian is given by
\begin{equation}
\mathcal{H}=\sum_{n}p_{n}\dot{q}_{n}-\mathcal{L}=\mathcal{H}_{0}%
+\mathcal{V},\nonumber
\end{equation}
where
\begin{equation}
\mathcal{H}_{0}=\frac{1}{2}\sum_{n}\left(  p_{n}^{2}+\omega_{n}^{2}q_{n}%
^{2}\right) \label{H_0}%
\end{equation}
and
\begin{align}
\mathcal{V}  & =-\frac{1}{2I_{c}^{2}}\sum_{n^{\prime},n^{\prime\prime
},n^{\prime\prime\prime},n^{\prime\prime\prime\prime}}p_{n^{\prime}%
}p_{n^{\prime\prime}}p_{n^{\prime\prime\prime}}p_{n^{\prime\prime\prime\prime
}}\\
& \times\int\limits_{0}^{l}dx\Delta Lu_{n^{\prime}}u_{n^{\prime\prime}%
}u_{n^{\prime\prime\prime}}u_{n^{\prime\prime\prime\prime}}.\nonumber
\end{align}
Quantization is achieved by regarding the variables $\left\{  q_{n}%
,p_{n}\right\}  $ as operators satisfying the following commutation relations
\begin{equation}
\left[  q_{n},p_{m}\right]  \equiv q_{n}p_{m}-p_{m}q_{n}=i\hbar\delta
_{n,m}\ \label{q,p}%
\end{equation}
and
\begin{equation}
\left[  q_{n},q_{m}\right]  =\left[  p_{n},p_{m}\right]
=0\ .\label{(q,q)(p,p)}%
\end{equation}
In terms of the Boson annihilation and creation operators
\begin{equation}
A_{n}=\frac{e^{i\omega_{n}t}}{\sqrt{2\hbar}}\left(  \sqrt{\omega_{n}}%
q_{n}+\frac{i}{\sqrt{\omega_{n}}}p_{n}\right)  ,\label{a}%
\end{equation}
\begin{equation}
A_{n}^{\dag}=\frac{e^{-i\omega_{n}t}}{\sqrt{2\hbar}}\left(  \sqrt{\omega_{n}%
}q_{n}-\frac{i}{\sqrt{\omega_{n}}}p_{n}\right)  ,\label{a+}%
\end{equation}
the Hamiltonian Eq.~(\ref{H_0}) can be expressed as
\begin{equation}
\mathcal{H}_{0}=\sum_{n}\hbar\omega_{n}\left(  A_{n}^{\dag}A_{n}+\frac{1}%
{2}\right)  .
\end{equation}
The current operator is given by
\begin{equation}
I\left(  x,t\right)  =\frac{\partial Q}{\partial t}=i\sum_{n}\sqrt{\frac
{\hbar\omega_{n}}{2}}\left(  A_{n}^{\dag}e^{i\omega_{n}t}-A_{n}e^{-i\omega
_{n}t}\right)  u_{n}\left(  x\right)
\end{equation}
and the voltage operator is given by
\begin{equation}
V\left(  x,t\right)  =-\frac{1}{C}\sum_{n}\sqrt{\frac{\hbar}{2\omega_{n}}%
}\left(  A_{n}^{\dag}e^{i\omega_{n}t}+A_{n}e^{-i\omega_{n}t}\right)
\frac{du_{n}}{dx}.
\end{equation}

The factors $p_{n^{\prime}}p_{n^{\prime\prime}}p_{n^{\prime\prime\prime}%
}p_{n^{\prime\prime\prime\prime}}$ generally contain terms oscillating rapidly
at frequencies on the order of the frequencies in the resonator spectrum. \ In
the RWA these terms are neglected, since their effect on the dynamics on a
time scale much longer than a typical oscillation period is negligibly small,
and only stationary terms remain \cite{Santamore 04 A}. \ Thus, in the
expression of $\mathcal{V}$ only terms of the type $p_{n^{\prime}}%
^{2}p_{n^{\prime\prime}}^{2}$ contain stationary terms which are given by
\begin{equation}
p_{n^{\prime}}^{2}p_{n^{\prime\prime}}^{2}\simeq\frac{\hbar\omega_{n^{\prime}%
}}{2}\frac{\hbar\omega_{n^{\prime\prime}}}{2}\left(  1+2A_{n^{\prime}}^{\dag
}A_{n^{\prime}}\right)  \left(  1+2A_{n^{\prime\prime}}^{\dag}A_{n^{\prime
\prime}}\right)  .
\end{equation}
The constant term can be disregarded, since it only gives rise to a constant
phase factor. \ Moreover, the terms $A_{n^{\prime}}^{\dag}A_{n^{\prime}}$ and
$A_{n^{\prime\prime}}^{\dag}A_{n^{\prime\prime}}$ that give rise to frequency
renormalization can be absorbed into $\mathcal{H}_{0}$. \ Thus, in the RWA the
perturbation $\mathcal{V}$ is given by
\begin{equation}
\mathcal{V}=\sum_{n^{\prime}\neq n^{\prime\prime}}\hbar\lambda_{n^{\prime
}n^{\prime\prime}}A_{n^{\prime}}^{\dag}A_{n^{\prime}}A_{n^{\prime\prime}%
}^{\dag}A_{n^{\prime\prime}}+\sum_{n^{\prime}}\hbar K_{n^{\prime}}\left(
A_{n^{\prime}}^{\dag}A_{n^{\prime}}\right)  ^{2},
\end{equation}
where
\begin{equation}
\lambda_{n^{\prime}n^{\prime\prime}}=-\frac{3}{I_{c}^{2}}\hbar\omega
_{n^{\prime}}\omega_{n^{\prime\prime}}\int\limits_{0}^{l}dx\Delta
Lu_{n^{\prime}}^{2}u_{n^{\prime\prime}}^{2}\label{lambda_n'n''}%
\end{equation}
and
\begin{equation}
K_{n^{\prime}}=-\frac{1}{2I_{c}^{2}}\hbar\omega_{n^{\prime}}^{2}%
\int\limits_{0}^{l}dx\Delta Lu_{n^{\prime}}^{4}.\label{K_n'}%
\end{equation}

\section{Perturbation Theory}

Consider the Schr\"{o}dinger equation
\begin{equation}
i\frac{d}{dt}\left\vert \psi\right\rangle =\mathcal{K}\left\vert
\psi\right\rangle ,
\end{equation}
where $\mathcal{K=K}^{\dag}$ is given by
\begin{equation}
\mathcal{K=K}_{0}+\lambda\mathcal{K}_{1},
\end{equation}
where $\lambda<<1$ is real. \ The time evolution operator $u\left(
t,t_{0}\right)  $ can be expanded as \cite{Weissbluth}
\begin{equation}
u\left(  t,t_{0}\right)  =u_{0}\left(  t,t_{0}\right)  +\mathcal{\lambda}%
u_{1}\left(  t,t_{0}\right)  +\mathcal{\lambda}^{2}u_{2}\left(  t,t_{0}%
\right)  +O\left(  \lambda^{3}\right)  ,
\end{equation}
where
\begin{equation}
u_{1}\left(  t,t_{0}\right)  =-i%
%TCIMACRO{\tint \limits_{t_{0}}^{t}}%
%BeginExpansion
{\textstyle\int\limits_{t_{0}}^{t}}
%EndExpansion
dt^{\prime}u_{0}\left(  t,t^{\prime}\right)  \mathcal{K}_{1}\left(  t^{\prime
}\right)  u_{0}\left(  t^{\prime},t_{0}\right)
\end{equation}
and
\begin{align}
& u_{2}\left(  t,t_{0}\right) \\
& =-%
%TCIMACRO{\tint \limits_{t_{0}}^{t}}%
%BeginExpansion
{\textstyle\int\limits_{t_{0}}^{t}}
%EndExpansion
dt^{\prime}%
%TCIMACRO{\tint \limits_{t_{0}}^{t^{\prime}}}%
%BeginExpansion
{\textstyle\int\limits_{t_{0}}^{t^{\prime}}}
%EndExpansion
dt^{\prime\prime}u_{0}\left(  t,t^{\prime}\right)  \mathcal{K}_{1}\left(
t^{\prime}\right)  u_{0}\left(  t^{\prime},t^{\prime\prime}\right)
\mathcal{K}_{1}\left(  t^{\prime\prime}\right)  u_{0}\left(  t^{\prime\prime
},t_{0}\right)  .\nonumber
\end{align}
Using this expansion, one can calculate the operator $O\left(  t\right)
\equiv u_{0}^{\dag}\left(  t,t_{0}\right)  u\left(  t,t_{0}\right)  $ to
second order in $\mathcal{\lambda}$
\begin{align}
& O\left(  t\right) \\
& =1-i\mathcal{\lambda}%
%TCIMACRO{\tint \limits_{t_{0}}^{t}}%
%BeginExpansion
{\textstyle\int\limits_{t_{0}}^{t}}
%EndExpansion
dt^{\prime}u_{0}\left(  t_{0},t^{\prime}\right)  \mathcal{K}_{1}\left(
t^{\prime}\right)  u_{0}\left(  t^{\prime},t_{0}\right) \nonumber\\
& -\mathcal{\lambda}^{2}%
%TCIMACRO{\tint \limits_{t_{0}}^{t}}%
%BeginExpansion
{\textstyle\int\limits_{t_{0}}^{t}}
%EndExpansion
dt^{\prime}%
%TCIMACRO{\tint \limits_{t_{0}}^{t^{\prime}}}%
%BeginExpansion
{\textstyle\int\limits_{t_{0}}^{t^{\prime}}}
%EndExpansion
dt^{\prime\prime}u_{0}\left(  t_{0},t^{\prime}\right)  \mathcal{K}_{1}\left(
t^{\prime}\right)  u_{0}\left(  t^{\prime},t^{\prime\prime}\right)
\mathcal{K}_{1}\left(  t^{\prime\prime}\right)  u_{0}\left(  t^{\prime\prime
},t_{0}\right) \nonumber
\end{align}
or
\begin{equation}
O\left(  t\right)  =1-i\mathcal{\lambda}%
%TCIMACRO{\tint \limits_{t_{0}}^{t}}%
%BeginExpansion
{\textstyle\int\limits_{t_{0}}^{t}}
%EndExpansion
dt^{\prime}\mathcal{K}_{1H}\left(  t^{\prime}\right)  -\mathcal{\lambda}^{2}%
%TCIMACRO{\tint \limits_{t_{0}}^{t}}%
%BeginExpansion
{\textstyle\int\limits_{t_{0}}^{t}}
%EndExpansion
dt^{\prime}%
%TCIMACRO{\tint \limits_{t_{0}}^{t^{\prime}}}%
%BeginExpansion
{\textstyle\int\limits_{t_{0}}^{t^{\prime}}}
%EndExpansion
dt^{\prime\prime}\mathcal{K}_{1H}\left(  t^{\prime}\right)  \mathcal{K}%
_{1H}\left(  t^{\prime\prime}\right)  ,
\end{equation}
where
\begin{equation}
\mathcal{K}_{1H}\left(  t\right)  \equiv u_{0}^{\dag}\left(  t,t_{0}\right)
\mathcal{K}_{1}\left(  t\right)  u_{0}\left(  t,t_{0}\right)  .
\end{equation}
Since $\mathcal{K}_{1}\left(  t\right)  $ is Hermitian, one finds to lowest
order in $\mathcal{\lambda}$
\begin{align}
& \left\vert \left\langle O\left(  t\right)  \right\rangle \right\vert ^{2}\\
& =1-\mathcal{\lambda}^{2}%
%TCIMACRO{\tint \limits_{t_{0}}^{t}}%
%BeginExpansion
{\textstyle\int\limits_{t_{0}}^{t}}
%EndExpansion
dt^{\prime}%
%TCIMACRO{\tint \limits_{t_{0}}^{t}}%
%BeginExpansion
{\textstyle\int\limits_{t_{0}}^{t}}
%EndExpansion
dt^{\prime\prime}\left[  \left\langle \mathcal{K}_{1}\left(  t^{\prime
}\right)  \mathcal{K}_{1}\left(  t^{\prime\prime}\right)  \right\rangle
-\left\langle \mathcal{K}_{1}\left(  t^{\prime}\right)  \right\rangle
\left\langle \mathcal{K}_{1}\left(  t^{\prime\prime}\right)  \right\rangle
\right] \nonumber
\end{align}
or
\begin{equation}
\left\vert \left\langle O\left(  t\right)  \right\rangle \right\vert
^{2}=1-\mathcal{\lambda}^{2}%
%TCIMACRO{\tint \limits_{t_{0}}^{t}}%
%BeginExpansion
{\textstyle\int\limits_{t_{0}}^{t}}
%EndExpansion
dt^{\prime}%
%TCIMACRO{\tint \limits_{t_{0}}^{t}}%
%BeginExpansion
{\textstyle\int\limits_{t_{0}}^{t}}
%EndExpansion
dt^{\prime\prime}\left\langle \mathcal{\tilde{K}}_{1}\left(  t^{\prime
}\right)  \mathcal{\tilde{K}}_{1}\left(  t^{\prime\prime}\right)
\right\rangle ,\label{|<O(t)>|^2}%
\end{equation}
where
\begin{equation}
\mathcal{\tilde{K}}_{1}\left(  t\right)  =\mathcal{K}_{1}\left(  t\right)
-\left\langle \mathcal{K}_{1}\left(  t\right)  \right\rangle .
\end{equation}

%use section* for acknowledgement

\section*{Acknowledgment}

%optional entry into table of contents (if used)
%\addcontentsline{toc}{section}{Acknowledgment}
E.B. would like to thank D. H. Santamore for fruitful discussions. \ This work
was supported by the German Israel Foundation under grant 1-2038.1114.07, the
Israel Science Foundation under grant 1380021, the Deborah Foundation, and
Poznanski Foundation.

\newpage
%Just because of unusual number of tables stacked at end
\bibliographystyle{plain}
\bibliography{apssamp}
%Produces the bibliography via BibTeX.

\end{document}